\documentclass[aps,prl,twocolumn,amsmath,amssymb,superscriptaddress,floatfix]{revtex4-1}
\usepackage{graphicx}
\usepackage{bm}
\usepackage{hyperref}
\usepackage{braket,bbold,color}

\newcommand{\bs}[1]{\boldsymbol{#1}}

\begin{document}

\title{Pairing Obstructions in Topological Superconductors}

\author{Frank Schindler}
\affiliation{Department of Physics, University of Zurich, Winterthurerstrasse 190, 8057 Zurich, Switzerland}

\author{Barry Bradlyn}
\affiliation{Department of Physics and Institute for Condensed Matter Theory, University of Illinois at Urbana-Champaign, Urbana, IL, 61801-3080, USA}

\author{Mark H. Fischer}
\affiliation{Department of Physics, University of Zurich, Winterthurerstrasse 190, 8057 Zurich, Switzerland}

\author{Titus Neupert}
\affiliation{Department of Physics, University of Zurich, Winterthurerstrasse 190, 8057 Zurich, Switzerland}

\begin{abstract}
The modern understanding of topological insulators is based on Wannier obstructions in position space. Motivated by this insight, we study topological superconductors from a position-space perspective. For a one-dimensional superconductor, we show that the wave function of an individual Cooper pair decays exponentially with separation in the trivial phase and polynomially in the topological phase.
For the position-space Majorana representation, we show that the topological phase is characterized by a nonzero Majorana polarization, which captures an irremovable and quantized separation of Majorana Wannier centers from the atomic positions. We apply our results to diagnose second-order topological superconducting phases in two dimensions.
Our work establishes a vantage point for the generalization of Topological Quantum Chemistry to superconductivity.
\end{abstract}

\maketitle


Topological phases of matter represent one of the main driving forces in condensed matter physics. In their most experimentally relevant realization, topological insulators~\cite{Kane05b,Bernevig06,Kane07,Koenig07,Hsieh08} (TIs) have garnered widespread attention owing to their protected surface states. 
These states can be predicted from the bulk electronic structure alone using topological invariants. All topological invariants discovered so far for translationally invariant, non-interacting systems can be phrased in terms of global momentum space properties~\cite{z2spinpumpfukane,Fu07,Fu11,BABHughesBenalcazar17,Brouwer17,SchindlerHOTI,Benalcazar17,Khalaf18,Geier18,BenMoTe2,wang2019two,SchindlerCorners}.

A modern approach to TIs is to characterize them in terms of their position-space structure~\cite{Slager12,Slager17,ashvin230indicators,Bradlyn17,Cano17,Cano17-2,BarryFragile,ZakEBR1,ZakEBR2,Alexandradinata18SureFire}. This point of view is informed by the insight that, if TIs are distinguished from trivial insulators by global properties in momentum space, then Heisenberg’s uncertainty principle suggests to study their local properties in position space. Within Topological Quantum Chemistry~\cite{Bradlyn17}, such a position-space approach has led to a classification and materials prediction program for TIs. In this approach, topology is defined as an obstruction to finding a gauge in which the Fourier transforms of Bloch eigenstates, called Wannier functions, are exponentially localized and preserve all of the symmetries.

\begin{figure}[t]
\centering
\includegraphics[width=.5\textwidth]{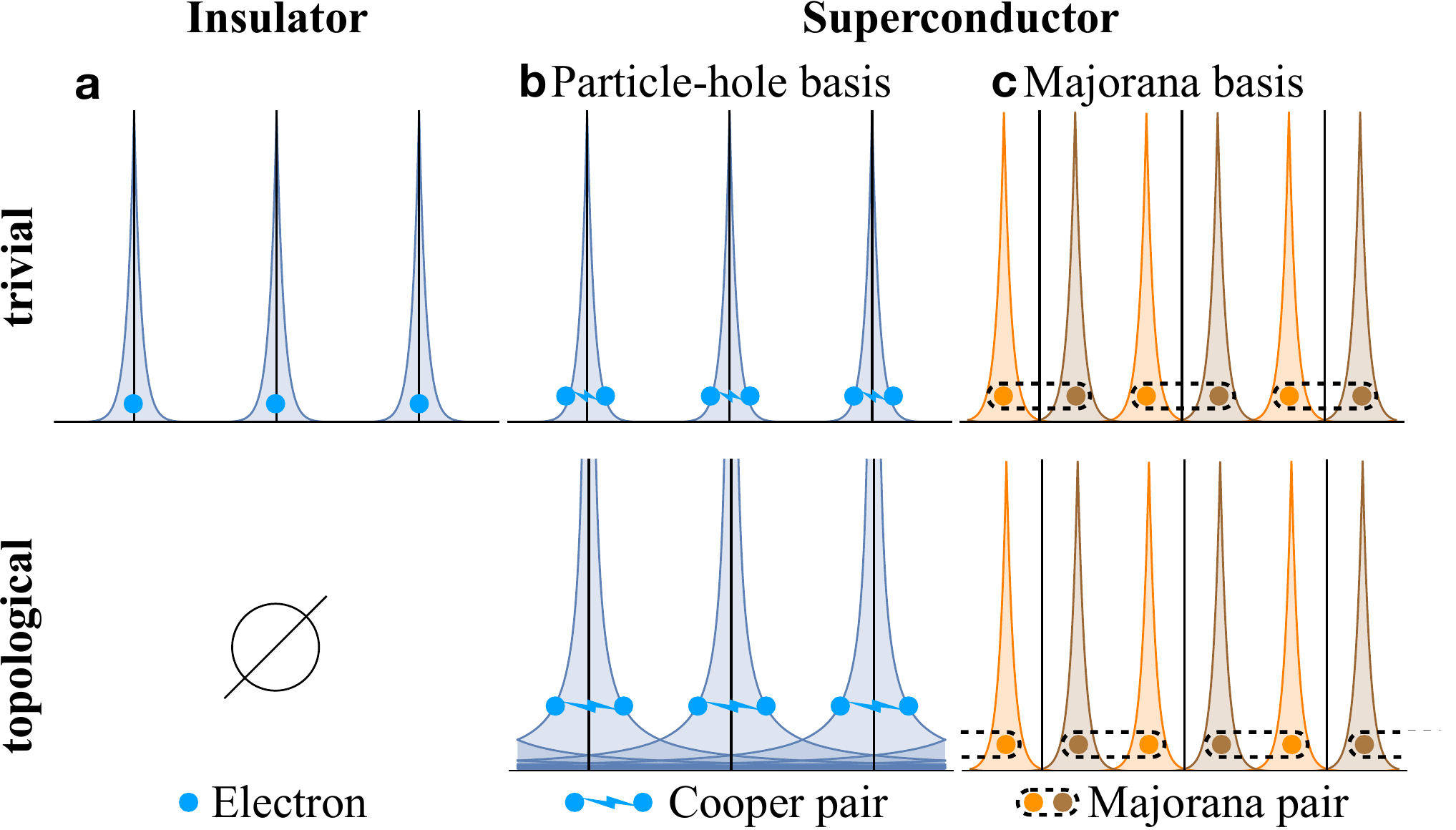}
\caption{Position-space picture of insulators and superconductors in 1D, with atomic sites indicated by vertical lines. 
\textbf{a}~The ground state of an insulator as a product over Wannier states. All 1D electronic band structures (absent symmetries) can be realized by exponentially localized Wannier functions. In contrast to superconductors, there are therefore no topological insulators in 1D.
\textbf{b}~The fixed-particle number component of the superconducting ground state as a product of Cooper pair states with wavefunctions decaying exponentially in the trivial phase and polynomially in the topological phase. 
\textbf{c}~The ground state as a product over Majorana Wannier functions with quantized \emph{Majorana charge}. In the topological phase, Majoranas from different atomic sites are paired up, leaving behind unpaired Majorana zero modes at the boundaries.}
\label{fig: 1Dresults}
\end{figure}

Another important class of topological phases of matter are topological superconductors (TSCs), which have a gapped bulk spectrum and exotic edge excitations that are their own anti-particles (so called Majorana modes)~\cite{Read00,Fu08,TanakaOddFrequency2012,Ando15,TanakaOddFrequency2019}. Recently, symmetry indicator invariants have been introduced for a momentum space based characterization of TSCs~\cite{Ono19,skurativska2019atomic,shiozaki2019variants,ono2019refined,geier2019symmetrybased}. In contrast to TIs, there is, however, as of yet no unifying physical picture of the position-space structure of TSCs.

In this letter, we present such a position-space picture of TSCs (focussing on class D in the Altland–Zirnbauer classification). The mean-field description of superconductors is formally equivalent to the band theory of non-interacting electrons. However, the ground state of a superconductor involves pairing of electrons and is therefore qualitatively different from the ground state of an insulator. This fact leads to striking differences between TSCs and TIs. In particular, we show that topology in superconductors can be phrased in terms of an obstruction to finding a gauge in which electron pairs are tightly (exponentially) bound. Furthermore, expressing the ground state in a basis of Majorana fermions, we find that the total spectral weight on the two Majorana degrees of freedom corresponding to each electron is split evenly, allowing us to introduce the concept of a quantized Majorana polarization. Our results are partially summarized in Fig.~\ref{fig: 1Dresults}.


\emph{1D $p$-wave superconductor in particle-hole basis---} We consider a 1D $p$-wave TSC, which could be realized by a nanowire with a single conduction band that is brought in proximity with an $s$-wave (trivial) superconductor~\cite{Roman10,Oreg10}. Within mean-field theory, the nanowire is described by the Hamiltonian $H = \sum_{k} \Psi^{\dagger}_{k} \mathcal{H}_{k} \Psi_{k}$, where we have introduced the Nambu spinor $\Psi_{k} = (c_{k}, c^\dagger_{-k})/\sqrt{2}$ and the Bogoliubov–de-Gennes (BdG) Hamiltonian
\begin{equation}
\label{eq: phconvention}
\mathcal{H}_{k} = \begin{pmatrix}
\epsilon_k & \Delta_k \\
\bar{\Delta}_k & -\epsilon_{-k}
\end{pmatrix},
\end{equation}
with $\mu$ the chemical potential and $\Delta_k = -\Delta_{-k}$ due to Fermi statistics (the bar denotes complex conjugation). The operator $c^\dagger_k$ creates an electron at momentum $k \in \{1, \dots, N\} 2\pi/N$, and $N$ is the number of sites (periodic boundary conditions are assumed). For $|\mu|<2|t|$, $H$ is in the topological phase, which hosts a single Majorana zero mode at each end of the nanowire, while for $|\mu|>2|t|$ the system is in the trivial phase. We refer to the convention in Eq.~\eqref{eq: phconvention} as the particle-hole basis. In contrast to insulators, the mean-field description of superconductors allows for the additional freedom of choosing a basis in Nambu space; this freedom is important for our position-space interpretation.
We can diagonalize this Hamiltonian by introducing Bogoliubov quasiparticle operators
\begin{equation}
\label{eq: BogoliubovDiagonalization}
\begin{aligned}
\begin{pmatrix} \alpha_{k} \\ \alpha^\dagger_{-k} \end{pmatrix} &= \begin{pmatrix} u_{k} & -v_{k} \\ -\bar{v}_{-k} & \bar{u}_{-k} \end{pmatrix} \begin{pmatrix} c_{k} \\ c^\dagger_{-k} \end{pmatrix} \equiv D_k \begin{pmatrix} c_{k} \\ c^\dagger_{-k} \end{pmatrix},
\end{aligned}
\end{equation}
where the matrix $D_k$ is unitary, with
\begin{equation}
\label{eq: constraintsduetounitarity}
\begin{aligned}
&u_{k} \bar{u}_{k} + v_{k} \bar{v}_{k} = u_{k} \bar{u}_{k} + v_{-k} \bar{v}_{-k} = \mathbb{1}, \\
&u_{k} v_{-k} + v_{k} u_{-k} = \bar{u}_{k} v_{k} + v_{-k} \bar{u}_{-k} = 0.
\end{aligned}
\end{equation}
For $|\mu|\neq2|t|$ the spectrum is gapped and $H$ has the ground state
\begin{equation}
\label{eq: manybodygroundstate}
\begin{aligned}
\ket{\Omega} & = \frac{1}{\mathcal{N}} \exp\left(\sum_{k} \frac{v_k}{u_k} c^\dagger_{k} c^\dagger_{-k} \right) \ket{0}, \\
& = \frac{1}{\mathcal{N}} \exp\left(\sum_{xy} g_{xy} c^\dagger_{x} c^\dagger_{y} \right) \ket{0},\\
& \hphantom{=} g_{xy} = \frac{1}{N} \sum_{k} e^{\mathrm{i} k (x-y)} \frac{v_k}{u_k},
\end{aligned}
\end{equation}
where $\ket{0}$ is the fermionic vacuum, $(-v_{-k},u_{-k})$ is the negative-energy eigenvector of the BdG Hamiltonian $\mathcal{H}_{k}$, and $\mathcal{N}$ is a normalization factor. In the second line, we introduced the Fourier-transformed operators 
\begin{equation}
\label{eq: fourierconvention}
c^\dagger_r = \frac{1}{N} \sum_{k} e^{-\mathrm{i} k r} c^\dagger_k, \quad r = 1 \dots N,
\end{equation}
that create a particle at site $r$ of the nanowire. We emphasize that the possibility of representating the ground state in the form of Eq.~\eqref{eq: manybodygroundstate}, reminiscent of a coherent state of Cooper pairs $c^\dagger_{k} c^\dagger_{-k}$, singles out the particle-hole basis, since it crucially relies on the anticommutation relations $\{c_{p},c^\dagger_{q}\} = \delta_{pq}, \{c_{p},c_{q}\} = \{c^{\dagger}_{p},c^{\dagger}_{q}\} = 0$ of electronic creation and annihilation operators.
Furthermore, extracting from Eq.~\eqref{eq: manybodygroundstate} the contribution to the $N$-particle state $\ket{\Omega_N}$, we obtain the amplitudes
\begin{equation}
\label{eq: superconductoratfixedparticlenumber}
\braket{r_1 \dots r_N |\Omega_N} \propto A\left[g_{r_1 r_2} g_{r_3 r_4} \dots g_{r_{N-1} r_N}\right],
\end{equation}
where $A[\cdot]$ denotes an antisymmetrization over all positions $r_1 \dots r_N$.
Invoking the Paley–Wiener theorem allows to determine the large-separation dependence of $g_{xy}$ on general grounds. In fact, $g_{xy}$ will fall off exponentially as $|x-y| \rightarrow \infty$ if the momentum space function $v_k/u_k$ is analytic~\cite{Strinati78, DubailRead15}. On the other hand, if $v_k/u_k$ diverges at some $k$, $g_{xy}$ will at most fall off polynomially with separation.

We now relate the analytical properties of $v_k/u_k$ to the topological characterization of the superconducting phase. We assume no further symmetries other than particle-hole symmetry (PHS), which in the particle-hole basis reads
\begin{equation}
\label{eq: phsdefinition}
P \mathcal{H}_{k} P^\dagger = -\mathcal{H}_{-k}, \quad P = \begin{pmatrix} 0 & \mathbb{1} \\ \mathbb{1} & 0 \end{pmatrix} \mathit{K}.
\end{equation}
PHS quantizes the Berry phase topological invariant
\begin{equation}
\label{eq: berryphasefor1Donebandpwave}
\begin{aligned}
\gamma &= \mathrm{i} \int \mathrm{d}k \, \left(\bar{u}_k \partial_k u_k + \bar{v}_k \partial_k v_k \right) \mod 2\pi \\
&= \frac{\mathrm{i}}{2} \int \mathrm{d}k \, \mathrm{Tr} \left[D^\dagger_k \partial_k D_k \right] \mod 2\pi \\
&= \int_{0}^{\pi} \mathrm{d}k \, \partial_k \lambda_k \mod 2\pi = (\lambda_\pi - \lambda_0) \mod 2\pi,
\end{aligned}
\end{equation}
where we wrote $\det D_k = e^{-\mathrm{i}\lambda_k}$ and used that PHS implies 
$
\lambda_{-k} = - \lambda_k \mod 2\pi,
$
and hence, $\lambda_0, \lambda_\pi = 0,\pi$.
For $\gamma$ to assume values $0,\pi$, we take the BdG functions $u_{k}$ and $v_{k}$ to be periodic functions in momentum space. This property, together with Eq.~\eqref{eq: fourierconvention}, corresponds to a convention where all atomic orbitals are located at the origin of the unit cell in position space. Our results can be straightforwardly generalized to arbitrary atomic positions~\footnote{See Supplemental Material for details, which includes Refs.~\cite{Read00,CoherentReview,Langbehn17,wang2018weak,SOTSC_Zhu19,SchindlerCorners,benalcazar2018quantization,Soluyanov2011Wannier,SoluyanovSmoothGauge,Song17,Alexandradinata14,BarryFragile,Taherinejad14}.}.

Now, if $\gamma = 0$, we can adiabatically deform the system to one in which only $u_k$ is nonzero and constant. Its inverse is then always well defined, and the ground state in position space, as expressed via $g_{xy}$, is a coherent superposition of exponentially-closely bound Cooper pairs. 
We next show that if $\gamma = \pi$, there are necessarily divergences in $v_k/u_k$, leading to a polynomial decay of $g_{xy}$. The proof proceeds by contradiction. Without loss of generality, we take $\gamma = \pi$ to be realized by $\lambda_0 = 0, \lambda_\pi = \pi$, implying $\det D_0 = 1$ and $\det D_\pi = -1$. Let us assume that $u_k$ is nonzero throughout momentum space. It therefore has a well-defined inverse, and we may reexpress $\det D_k$ as
\begin{equation}
\begin{aligned}
\det D_k &=  u_k \bar{u}_{-k} \left(\frac{u_k \bar{u}_{-k}-v_k \bar{v}_{-k}}{u_k \bar{u}_{-k}}\right) \\
&= u_k \bar{u}_{-k} \left(1 - \frac{v_k \bar{v}_{-k}}{u_k \bar{u}_{-k}}  \right) \\
&= \frac{u_k}{u_{-k}} = 1 \text{ at } k=0,\pi,
\end{aligned}
\end{equation}
where we used the constraints in Eq.~\eqref{eq: constraintsduetounitarity}. This result is, however, in contradiction to our earlier assertion that $\det D_\pi = -1$. We conclude that either $u_0$ or $u_\pi$ are zero in systems with $\gamma = \pi$. This result carries over to an arbitrary number of bands~\cite{Note1}.

Thus, the long-distance behavior of the Cooper pair wavefunction is indicative of the topological character of the phase: In the trivial phase, there is strong pairing and the wavefunction decays exponentially with separation, while in the topological case, there is weak pairing and the wavefunction decays only polynomially. This \emph{pairing obstruction} is in close correspondence to the Wannier obstruction of TIs~\cite{Brouder2007Exponential,Soluyanov2011Wannier,Bradlyn17}, but in contrast to insulators~\footnote{In this work, we define insulators as spinful fermionic systems that have gapped and charge-preserving ground states, and therefore belong to the Altland-Zirnbauer classes A or AII.}, the particle-hole symmetry of superconductors allows for topological phases already in 1D.

\begin{figure}[t]
\centering
\includegraphics[width=.5\textwidth]{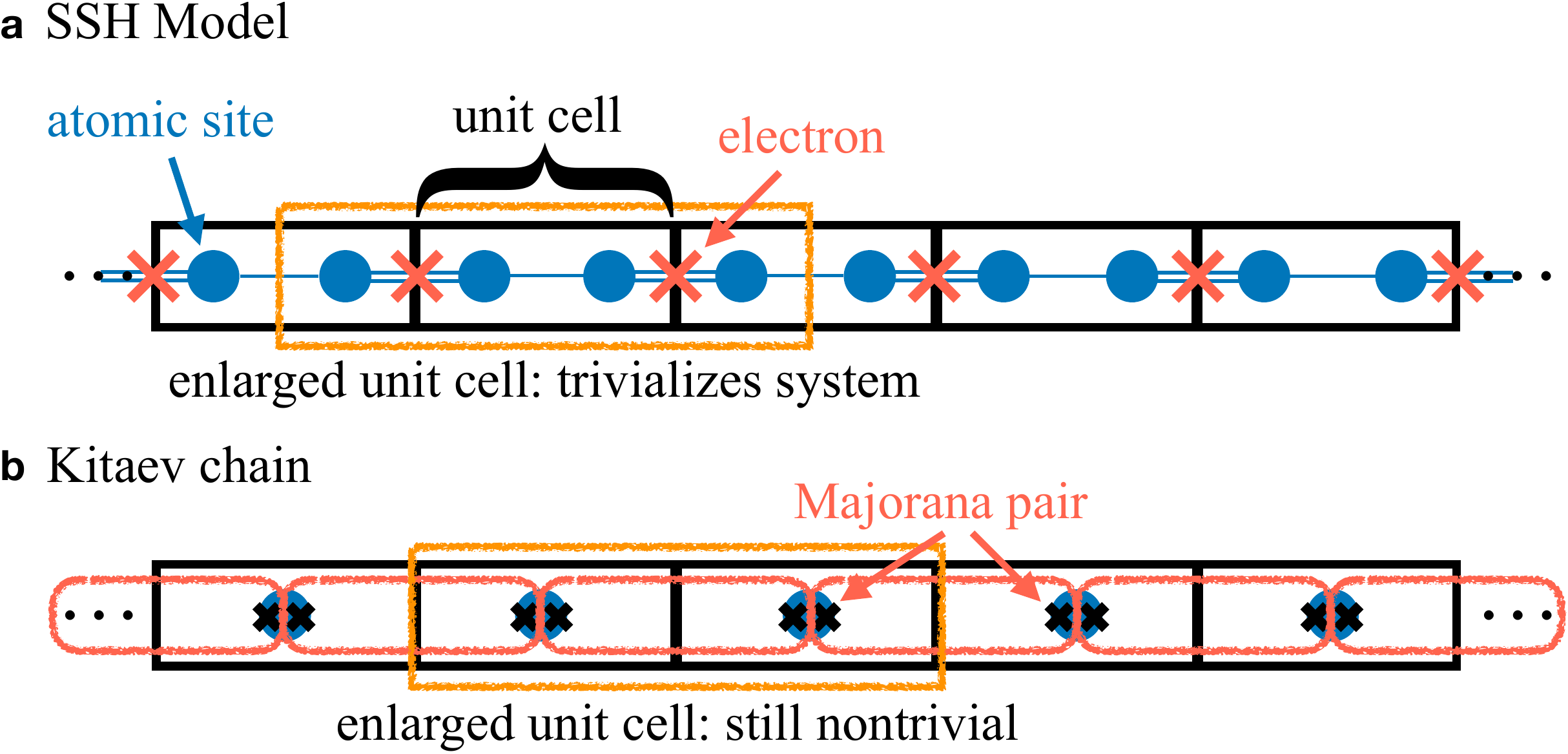}
\caption{Comparison of 1D topological insulators and superconductors. \textbf{a}~In the Su-Schrieffer-Heeger (SSH) model~\cite{SSHoriginal}, spatial inversion symmetry is required to pin electrons to the unit cell edges, leading to midgap states in an open geometry. However, it is possible to choose an inversion-symmetric enlarged unit cell that has no end states. \textbf{b}~In the Kitaev chain~\cite{Kitaev_2001}, Majorana modes are paired across unit cells, leaving behind unpaired Majorana zero modes in an open geometry. Since unit cells may not cut through atomic sites, any enlarged unit cell shows the same topological behavior.}
\label{fig: SSHvsKitaev}
\end{figure}

\emph{1D $p$-wave superconductor in Majorana basis---}
We introduce the Majorana modes $a_k = c_{k} + c^\dagger_{-k}$, $b_k = (c_{k} - c^\dagger_{-k})/\mathrm{i}$, and the Majorana BdG Bloch functions $v^\mathrm{M}_{k} = (v_{k} - u_{k})/\sqrt{2}$ and $u^\mathrm{M}_{k} = \mathrm{i}(v_{k} + u_{k})/\sqrt{2}$. Let $a_r, b_r$ be the Fourier transforms of the Majorana modes [using the same convention as in Eq.~\eqref{eq: fourierconvention}]. We can then reexpress the ground state in Eq.~\eqref{eq: manybodygroundstate} in terms of Majorana Wannier functions $W^{\mathrm{M}}_{R \alpha}(r)$, $\alpha = a,b$, which we define by
\begin{equation}
\label{eq: MajoranaWannierGroundState}
\begin{aligned}
&\ket{\Omega} = \frac{1}{\mathcal{N}} \prod_{R} \left(\sum_{r} W^{\mathrm{M}}_{R a}(r) a_r + W^{\mathrm{M}}_{R b}(r) b_r \right) \ket{0},\\
&\begin{pmatrix} W^{\mathrm{M}}_{R a}(r) \\ W^{\mathrm{M}}_{R b}(r) \end{pmatrix} = \frac{1}{N} \sum_{k} e^{- \mathrm{i} k (R-r)} \begin{pmatrix} -v^\mathrm{M}_k \\ u^\mathrm{M}_k \end{pmatrix},
\end{aligned}
\end{equation}
where $R$ labels the position of the unit cell that the Wannier functions belong to.
As for any 1D system~\cite{VanderbiltReview}, the Wannier functions $W^{\mathrm{M}}_{R \alpha}(r)$ can be exponentially localized. In the unit cell, they are centered around the position 
\begin{equation}
\label{eq: majoranawanniercenterdefinition}
x_\mathrm{o} = \sum_{r,\alpha} \overline{W}^{\mathrm{M}}_{0 \alpha}(r) r W^{\mathrm{M}}_{0 \alpha}(r) \mod 1.
\end{equation}
(The subscript $\mathrm{o}$ stands for the occupied subspace spanned by the Majorana Wannier functions.)
Particle-hole symmetry implies that $x_\mathrm{o} = x_\mathrm{e}$, where $x_\mathrm{e}$ denotes the center of Wannier functions built from the empty Majorana bands. The Wannier centers for all bands (which form a complete basis for the Hilbert space) are always exponentially localized on the lattice sites. We therefore have $x_\mathrm{o} + x_\mathrm{e}  \mod 1 = 0$. We conclude that $x_\mathrm{o} = 0, 1/2$ is quantized and provides a topological invariant characterizing the many-body ground state.
In fact, we know from the general theory of maximally localized Wannier functions~\cite{VanderbiltReview} that
$
x_\mathrm{o} = \gamma/2\pi = 0, 1/2,
$
where $\gamma$ is defined similar to Eq.~\eqref{eq: berryphasefor1Donebandpwave}:
\begin{equation}
\label{eq: berryphasefor1DMajorana}
\gamma = \mathrm{i} \int \mathrm{d}k \, \left(\bar{u}^\mathrm{M}_k \partial_k u^\mathrm{M}_k + \bar{v}^\mathrm{M}_k \partial_k v^\mathrm{M}_k \right) \mod 2\pi
\end{equation}
(the two definitions are equivalent).
Therefore, the Wannier states in Eq.~\eqref{eq: MajoranaWannierGroundState} can be adiabatically continued back to the original delta-localized Majorana basis states at position $R$, created by the bare operators $a_R$ and $b_R$, if and only if the Berry phase vanishes: These original basis states have Wannier functions $W^{\mathrm{M}}_{R \alpha}(r) = \delta_{R,r} \delta_{\alpha,1}$ or $W^{\mathrm{M}}_{R \alpha}(r) = \delta_{R,r} \delta_{\alpha,2}$, both of which correspond to $x_\mathrm{o} = 0$. This property implies a Majorana pairing obstruction: It is impossible to adiabatically connect the topological superconducting ground state to a collection of physical (i.e., deriving from the atomic positions) electrons or holes. In contrast, in a trivial superconductor it is possible to turn off superconductivity without closing the gap between positive and negative-energy quasiparticle states.

We next show that the Majorana representation is set apart from other basis decompositions of $H$ in that it allows for a meaningful generalization of polarization to superconductors. Taking the trace of the Majorana version of the constraint in Eq.~\eqref{eq: constraintsduetounitarity}, we find that
\begin{equation}
\begin{aligned}
&\sum_{k} \bar{v}^\mathrm{M}_{k} v^\mathrm{M}_{k} = \frac{N}{2} = \sum_{k} \bar{u}^\mathrm{M}_{-k} u^\mathrm{M}_{-k}, \\
&\sum_{r} \overline{W}^{\mathrm{M}}_{0 \alpha}(r) W^{\mathrm{M}}_{0 \alpha}(r) = \frac{1}{N} \sum_{k} \begin{pmatrix} \bar{v}^\mathrm{M}_{k} v^\mathrm{M}_{k} \\ \bar{u}^\mathrm{M}_{k} u^\mathrm{M}_{k} \end{pmatrix}_\alpha = \frac{1}{2}.
\end{aligned}
\end{equation}
This implies that the total spectral weight carried by Majoranas of $a$ or $b$ type is always equal, a property that is not realized in other bases: In the particle-hole basis, for example, the total spectral weight carried by holes is zero in the case of a band insulator. The total Wannier function support within a unit cell splits into equal contributions of Majoranas of $a$ and $b$ type. This result carries over to an arbitrary number of bands~\cite{Note1}.

We therefore introduce a quantized \emph{Majorana charge}, nominally equal to $1/2$, and a corresponding Majorana polarization~\footnote{Our concept of Majorana polarization differs from that of Ref.~\onlinecite{Simon12}. The latter work introduced a continuous and local quantity that captures the Majorana character of a given state in Nambu space, similar to how spin polarization captures local spin alignment. The Majorana polarization defined here is a quantized and global bulk property, and guarantees the presence of Majorana end states.} that is computed via Eq.~\eqref{eq: majoranawanniercenterdefinition}, or alternatively via the Berry phase in Eq.~\eqref{eq: berryphasefor1DMajorana}. Figure~\ref{fig: SSHvsKitaev} shows how Majorana polarization, unlike electronic polarization, survives translational symmetry breaking. We can now see how the anomalous end states of a 1D superconductor arise in the topological regime, because a Majorana polarization of $1/2$ results in a single Majorana mode localized at the end of an open geometry. Due to particle-hole symmetry, this mode is necessarily a zero-energy state.

\begin{table}[]
\begin{tabular}{|l|l|l|l|}
\hline
dimension              & $g_{xy}$ decay& $W^{\mathrm{M}}_{R \alpha}(r)$ decay& phase label \\ \hline \hline
1D      & exponential          & exponential             & trivial             \\ 
           & polynomial          & exponential             & $p$-wave TSC             \\ \hline \hline
2D      & exponential         & exponential            & trivial             \\
          & polynomial          & polynomial             & chiral TSC             \\ 
   	 & polynomial          & exponential             & 2nd order TSC            \\ \hline \hline
3D      & exponential          & exponential             & trivial             \\ 
	& polynomial          & polynomial             & 2nd order TSC             \\ 
	& polynomial          & exponential             & 3rd order TSC            \\ \hline
\end{tabular}
\caption{\label{tab: overview} Overview of topological superconductors without time-reversal symmetry. The 1D $p$-wave TSC hosts zero-dimensional Majorana end states. The 2D chiral TSC can be obtained from it via a Thouless pump, which naturally explains why its Wannier functions cannot be exponentially localized. Another possibility in 2D is the 2nd order TSC with Majorana corner states. In 3D, the only Wannier obstructed superconductor absent time-reversal symmetry is the 2nd order TSC with 1D Majorana hinge states similar to the edge states of the 2D $p$-wave TSC.}
\end{table}

\emph{2D chiral superconductor and generalization to higher dimensions---} The two-dimensional $p$-wave superconductor~\cite{Read00, KITAEV20062} can be obtained from a Thouless pump of bulk Majorana fermions.
Such a superconductor is characterized by a nonzero Chern number of the occupied BdG eigenstates. We interpret this Chern number as a flow of the Majorana separation from the atomic sites, Eq.~\eqref{eq: majoranawanniercenterdefinition}, with a transversal momentum coordinate.
Consider a two-dimensional BdG Hamiltonian $\mathcal{H}_{\bs{k}}$. Particle-hole symmetry implies $P \mathcal{H}_{\bs{k}} P^\dagger = -\mathcal{H}_{-\bs{k}}$, with $P$ defined in Eq.~\eqref{eq: phsdefinition}. Writing $\bs{k} = (k_x, k_y)$, there are two special values of $k_y = 0,\pi$, at which $\mathcal{H}_{(k_x,k_y = 0,\pi)} \equiv \mathcal{H}_{k_x}^{0,\pi}$ can be interpreted as the BdG Hamiltonian of a 1D superconductor.
In the Majorana basis, we introduce hybrid Wannier functions that are indexed by $k_y$, affording a $k_y$-dependent Majorana polarization
\begin{equation}
x^{k_y}_\mathrm{o} = \sum_{r,\alpha} \overline{W}^{\mathrm{M}}_{0 \alpha}(k_y, r_x) r_x W^{\mathrm{M}}_{0 \alpha}(k_y, r_x) \mod 1.
\end{equation}
We note again that we are working in a convention where the atomic positions are all at $\bs{r}_i = (0,0)$. Crucially, due to the action of particle-hole symmetry, the related Berry phase is only quantized for $k_y = 0,\pi$, namely
$x^{0,\pi}_\mathrm{o} = \gamma^{0,\pi}/2\pi = 0, 1/2,$
where $\gamma^{k_y}$ is evaluated along 1D momentum space slices of constant $k_y$.
The Berry phase winding as a function of $k_y$ is related to the Chern number
\begin{equation}
C = \frac{1}{2\pi} \int \mathrm{d}k_y \, \partial_{k_y}\gamma^{k_y} = \int \mathrm{d}k_y \, \partial_{k_y}x^{k_y}_\mathrm{o}.
\end{equation}
We conclude that in a 2D $p_x + \mathrm{i} p_y$ superconductor with $C = 1$, the Majorana polarization $x^{k_y}_\mathrm{o}$ continuously evolves from $0$ to $1 \equiv 0$ as $k_y$ undergoes a noncontractible cycle. Importantly, this implies that it is impossible to exponentially localize the Majorana Wannier functions $W^{\mathrm{M}}_{0 \alpha}(k_y, r_x)$ also in the $y$-direction, as they are not smooth functions of $k_y$.

We therefore find that a topological 2D $p$-wave superconductor admits neither an exponentially decaying Cooper pair function $g_{\bs{x}\bs{y}}$, nor exponentially decaying Majorana Wannier functions $W^{\mathrm{M}}_{\bs{R} \alpha}(\bs{r})$. A natural question is then if there exists a 2D superconductor with a polynomially decaying Cooper pair function but exponentially decaying Majorana Wannier functions. This ``Majorana obstructed atomic limit" phase is naturally realized by the recently discovered higher-order TSCs, which feature Majorana zero modes localized at the corners of samples terminated in both $x$ and $y$ directions~\cite{Brouwer17,Khalaf18,wang2018weak,hsu2018majorana} (Table~\ref{tab: overview} provides an overview). Our results imply that these phases are ``stronger" than their insulating counterparts, in that they cannot be trivialized by the introduction of uncharged ancillas~\cite{Else2019InteractingFragile}. 

\begin{figure}[t]
\centering
\includegraphics[width=.5\textwidth]{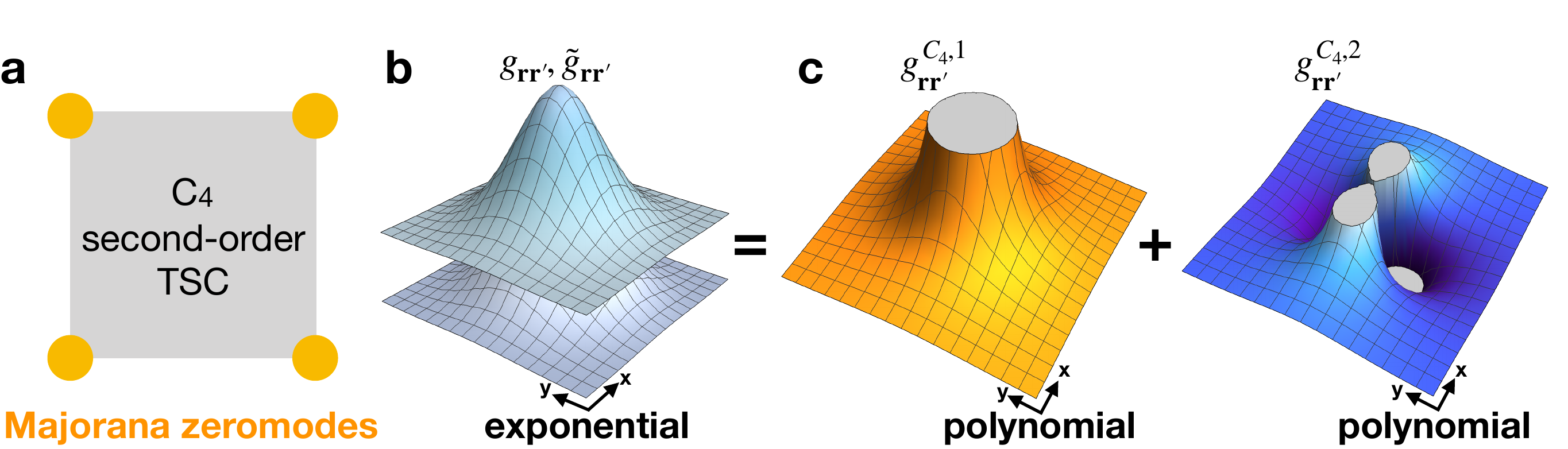}
\caption{ Real-space structure of a $C_4$-symmetric second-order topological superconductor. {\textbf a}~Majorana corner modes in a square-shaped sample geometry. {\textbf b}~Naively, the two ground state Cooper pair wavefunctions [generalizations of Eq.~\eqref{eq: manybodygroundstate}] $g_{\mathbf{r} \mathbf{r}'}$ and $\tilde{g}_{\mathbf{r} \mathbf{r}'}$ decay exponentially with electron separation. {\textbf c}~When chosen as eigenfunctions of $C_4$ symmetry, the Cooper pair wavefunctions of each $C_4$ subsector, $g^{C_4,1}_{\mathbf{r} \mathbf{r}'}$ and $g^{C_4,2}_{\mathbf{r} \mathbf{r}'}$, necessarily decay polynomially as a consequence of the nontrivial second-order topology.}
\label{fig: c4sotsc}
\end{figure}

Our methods are in particular insightful for topological superconductors that cannot be diagnosed by standard topological invariants such as (non-abelian) Berry phases and symmetry indicators. An example are second-order topological superconductors protected by $C_4$ rotation symmetry and spinful time-reversal~\cite{Song17,benalcazar2018quantization,SchindlerCorners}. These host four Kramers pairs of Majorana zero-energy states at the corners of a square-shaped sample, and otherwise have a gapped bulk and gapped edge spectrum. Generalizing our results to 2D systems with crystalline symmetries, they too can be identified by the decay behavior of their Cooper pair wavefunctions: we show in the Supplemental Material~\cite{Note1} that the Cooper pairs \emph{in individual $C_4$ eigenvalue subspaces} are subject to a pairing obstruction. See Fig.~\ref{fig: c4sotsc} for an illustration.

\emph{Discussion---} The mean-field BdG description of superconductors is uncannily similar to the Bloch description of non-interacting electrons. This has led to a remarkable amount of concept transfer and cross-fertilization between the descriptions of topological superconductors and insulators. However, it is often not clear what becomes of the physical interpretation of the mathematical quantities involved. An example is the Berry phase, which captures the electric polarization of an insulator, something that is evidently not well defined for superconductors. Employing a position-space picture phrased in terms of Cooper pairs and Majorana excitations, our work gives physical meaning to the Berry phase in the superconducting context via the notion of pairing obstruction and Majorana polarization. A natural next step would be a comprehensive inclusion of symmetries into our framework. Another area of potential future work is to use our formalism to understand the peculiarities of number-conserving models of topological superconductors~\cite{LeggettParticleConserving18,Knapp2019,Lapa2019,Lapa20}.

\begin{acknowledgments}
FS thanks Roman Lutchyn, Bela Bauer, and William Cole for helpful discussions. TN thanks Piet Brouwer for helpful discussions. BB and TN thank the Banff International Research Station for hosting during some stages of this work. FS was supported by the Swiss National Science Foundation (grant number: 200021\_169061), the National Science Foundation under Grant No. NSF PHY-1748958, and by the Heising-Simons Foundation. This project has received funding from the European Research Council under the European Union’s Horizon 2020 research and innovation program (ERC-StG-Neupert-757867-PARATOP).
\end{acknowledgments}

\bibliography{Ref-Lib}

\end{document}


\title{Pairing Obstructions in Topological Superconductors: \\Supplementary Information}

\author{Frank Schindler}
\affiliation{Department of Physics, University of Zurich, Winterthurerstrasse 190, 8057 Zurich, Switzerland}

\author{Barry Bradlyn}
\affiliation{Department of Physics and Institute for Condensed Matter Theory, University of Illinois at Urbana-Champaign, Urbana, IL, 61801-3080, USA}

\author{Mark H. Fischer}
\affiliation{Department of Physics, University of Zurich, Winterthurerstrasse 190, 8057 Zurich, Switzerland}

\author{Titus Neupert}
\affiliation{Department of Physics, University of Zurich, Winterthurerstrasse 190, 8057 Zurich, Switzerland}

\maketitle

\tableofcontents
\section{Formalism}
\label{sec: formalism}
We here introduce in detail the mean-field formalism of multi-band superconductivity in $D$ spatial dimensions.
\subsection{Particle-hole basis}
A translationally invariant superconducting electronic system with $M$ bands is described by a mean-field Hamiltonian of the form
\begin{equation}
\label{eq: BdGHamiltonian}
\begin{aligned}
H = \sum_{\bs{k} ij} \bigg[&h_{\bs{k} ij} c^\dagger_{\bs{k} i} c_{\bs{k} j} +
 \frac{1}{2} \left(\Delta_{\bs{k} ij} c^\dagger_{\bs{k} i} c^\dagger_{-\bs{k} j} + \Delta^{\dagger}_{\bs{k} ij} c_{-\bs{k} i} c_{\bs{k} j} \right)\bigg],
\end{aligned}
\end{equation}
where $k_n \in \{1, \dots, N\} 2\pi/N $, $n = 1 \dots D$, denotes the set of good momentum quantum numbers of a $D$-dimensional system of linear extent $N$ (we assume a hypercubic lattice for simplicity), and $i,j = 1 \dots M$ run over the degrees of freedom within each unit cell. The operator $c_{\bs{k} i}$ annihilates an electron of momentum $\bs{k}$ in orbital $i$ and we have the algebra
\begin{equation}
\label{eq: commutationalgebraforkspaceoperators}
\begin{aligned}
\{c_{\bs{k}i},c^\dagger_{\bs{q} j}\} &= \delta_{\bs{k},\bs{q}} \delta_{i,j},\\
\{c_{\bs{k}i},c_{\bs{q} j}\} &=
\{c^{\dagger}_{\bs{k}i},c^{\dagger}_{\bs{q} j}\} = 0,
\end{aligned}
\end{equation}
where $\{\cdot,\cdot\}$ denotes the anticommutator.
We can rewrite the Hamiltonian in Bogoliubov–de-Gennes (BdG) form, obtaining
\begin{equation}
\label{eq: BdGHamiltonianDecompositionInPHBasis}
H = \sum_{\bs{k} ij} \Psi^{\dagger}_{\bs{k} i} \mathcal{H}_{\bs{k} ij} \Psi_{\bs{k} j},
\end{equation}
where we have used matrix notation and introduced
\begin{equation}
\begin{aligned}
\Psi_{\bs{k} i} &= \frac{1}{\sqrt{2}} \begin{pmatrix} c_{\bs{k} i} \\ c^\dagger_{-\bs{k} i} \end{pmatrix}, \\
\mathcal{H}_{\bs{k} ij} &= \begin{pmatrix}
h_{\bs{k} ij} & \Delta_{\bs{k} ij} \\
\Delta^{\dagger}_{\bs{k} ij} & -h^{\mathrm{T}}_{-\bs{k} ij}
\end{pmatrix}.
\end{aligned}
\end{equation}
The fermionic statistics of the Cooper pairs implies 
\begin{equation}
\Delta_{\bs{k} ij}=-\Delta_{-\bs{k} ji}.
\end{equation}
The particle-hole symmetry $P$ reads
\begin{equation}
\label{eq: phsdefinition}
P = \begin{pmatrix} 0 & \mathbb{1} \\ \mathbb{1} & 0 \end{pmatrix} \mathit{K}, \quad P H P^\dagger = H,
\end{equation}
which imposes the constraint $P \mathcal{H}_{\bs{k}} P^\dagger = -\mathcal{H}_{-\bs{k}}$.
We may now diagonalize $\mathcal{H}_{\bs{k} ij}$ and write
\begin{equation}
\label{eq: diagBdGH}
\begin{aligned}
&\sum_{j} \mathcal{H}_{\bs{k} ij} \begin{pmatrix} \bar{u}_{\bs{k} \xi j} \\ -\bar{v}_{\bs{k} \xi j} \end{pmatrix} = E_{\bs{k} \xi} \begin{pmatrix} \bar{u}_{\bs{k} \xi i} \\ -\bar{v}_{\bs{k} \xi i} \end{pmatrix},\\
\xrightarrow[]{P} &\sum_{j} \mathcal{H}_{\bs{k} ij} \begin{pmatrix} -v_{-\bs{k} \xi j} \\ u_{-\bs{k} \xi j} \end{pmatrix} = -E_{-\bs{k} \xi} \begin{pmatrix} -v_{-\bs{k} \xi i} \\ u_{-\bs{k} \xi i} \end{pmatrix},
\end{aligned}
\end{equation}
where a bar denotes complex conjugation, $\xi = 1 \dots M$ labels the eigenstates that are not related to each other by particle-hole symmetry, and we chose $E_{\bs{k} \xi} > 0$. Eq.~\eqref{eq: diagBdGH} is the complex conjugate of the BdG equation in Ref.~\onlinecite{Read00}. Here and in the following, we take the BdG functions $u_{\bs{k} \xi j}$ and $v_{\bs{k} \xi j}$ to be periodic functions in the Brillouin zone. This corresponds to a convention where all atomic orbitals are located at the origin of the unit cell in real space. Importantly, the actual atomic positions are irrelevant for the salient features of our analysis, as we do not consider crystalline symmetries in this work. In Sec.~\ref{sec: nonperiodicbandappendix}, we nevertheless generalize our results to arbitrary atomic positions in order to make contact with the existing literature on crystalline topological insulators.

A spectral decomposition of $\mathcal{H}_{\bs{k} ij}$ yields
\begin{equation}
\label{eq: diagonalizedHamiltonian}
\begin{aligned}
H &= \sum_{\bs{k} \xi} E_{\bs{k} \xi} \alpha^\dagger_{\bs{k}\xi} \alpha_{\bs{k}\xi}, \\
\alpha_{\bs{k}\xi} &= \sum_{i} \left[ u_{\bs{k} \xi i} c_{\bs{k} i} - v_{\bs{k} \xi i} c^{\dagger}_{-\bs{k} i} \right].
\end{aligned}
\end{equation}
Choosing the eigenvectors of $\mathcal{H}_{\bs{k} ij}$ to be orthonormal implies that 
\begin{equation}
\begin{aligned}
\{\alpha_{\bs{k}\xi},\alpha^\dagger_{\bs{q} \zeta}\} &= \delta_{\bs{k},\bs{q}} \delta_{\xi,\zeta},\\
\{\alpha_{\bs{k}\xi},\alpha_{\bs{q} \zeta}\} &=
\{\alpha^{\dagger}_{\bs{k}\xi},\alpha^{\dagger}_{\bs{q} \zeta}\} = 0.
\end{aligned}
\end{equation}
This ensures that we can write the superconducting ground state $\ket{\Omega}$ of $H$, which satisfies $\alpha_{\bs{k}\xi} \ket{\Omega} = 0$, as
\begin{equation}
\label{eq: superconductingGSwrittenwithalphas}
\ket{\Omega} \propto \prod_{\bs{k},\xi} \alpha_{\bs{k}\xi} \ket{0},
\end{equation}
where the electronic vacuum $\ket{0}$ is defined by $c_{\bs{k} i} \ket{0} = 0$.
We note that the eigenstate decomposition in Eq.~\eqref{eq: diagonalizedHamiltonian} can be rewritten in matrix form:
\begin{equation}
\begin{pmatrix} \bs{\alpha}_{\bs{k}} \\ \bs{\alpha}^\dagger_{-\bs{k}} \end{pmatrix} = \begin{pmatrix} u_{\bs{k}} & -v_{\bs{k}} \\ -\bar{v}_{-\bs{k}} & \bar{u}_{-\bs{k}} \end{pmatrix} \begin{pmatrix} \bs{c}_{\bs{k}} \\ \bs{c}^\dagger_{-\bs{k}} \end{pmatrix},
\end{equation}
where $\bs{\alpha}_{\bs{k}}$ is a vector with elements $\alpha_{\bs{k}\xi}$, $\xi = 1 \dots M$, and $\bs{c}_{\bs{k}}$ is a vector with elements $c_{\bs{k} i}$, $i = 1 \dots M$. The condition that this map be unitary translates to
\begin{equation}
\label{eq: uudaggerconstraint}
u_{\bs{k}} u_{\bs{k}}^\dagger + v_{\bs{k}} v_{\bs{k}}^\dagger = u^\dagger_{\bs{k}} u_{\bs{k}} + \bar{v}^\dagger_{-\bs{k}} \bar{v}_{-\bs{k}} = \mathbb{1},
\end{equation}
\begin{equation}
\label{eq: uvtransposeconstraint}
u_{\bs{k}} \bar{v}_{-\bs{k}}^\dagger + v_{\bs{k}} \bar{u}_{-\bs{k}}^\dagger = u_{\bs{k}}^\dagger v_{\bs{k}} +\bar{v}^\dagger_{-\bs{k}} \bar{u}_{-\bs{k}} = 0.
\end{equation}
At a momentum $\bar{\bs{k}} = - \bar{\bs{k}}$ that is equal to its opposite up to a reciprocal lattice vector (there are $2^D$ such momenta), this implies 
\begin{equation}
\label{eq: pinningconstraintofdets}
\left[1-(-1)^M \right] \det u_{\bar{\bs{k}}} \det v_{\bar{\bs{k}}} = 0. 
\end{equation}
We can rewrite the ground state as a coherent state of Cooper pairs. Comparing with Eqs.~(6.36)-(6.62) of Ref.~\onlinecite{CoherentReview}, we find that the ground state is (up to a normalization factor) given by
\begin{equation}
\label{eq: manybodygroundstate}
\ket{\Omega} \propto \exp\left(\sum_{\bs{k} ij} g_{\bs{k} ij} c^\dagger_{\bs{k} i} c^\dagger_{-\bs{k} j} \right) \ket{0},\quad
g_{\bs{k}} = u^{-1}_{\bs{k}} v_{\bs{k}}.
\end{equation}

\subsection{Majorana basis}
We rewrite the fermionic creation and annihilation operators in terms of Majorana operators as
\begin{equation}
\begin{aligned}
c_{\bs{k} i} = \frac{1}{2} (a_{\bs{k} i} + \mathrm{i} b_{\bs{k} i})&, \quad c^\dagger_{\bs{k} i} = \frac{1}{2} (a_{-\bs{k} i} - \mathrm{i} b_{-\bs{k} i}) \\
a_{\bs{k} i} = c_{\bs{k} i} + c^\dagger_{-\bs{k} i}&, \quad b_{\bs{k} i} = \frac{c_{\bs{k} i} - c^\dagger_{-\bs{k} i}}{\mathrm{i}}.
\end{aligned}
\end{equation}
We see that the Majorana operators satisfy 
\begin{equation}
\label{eq: Majoranacommutators}
\begin{aligned}
&a_{\bs{k} i}^2 = b_{\bs{k} i}^2 = \delta_{\bs{k},-\bs{k}}, \\ &\{a_{\bs{k} i},a_{\bs{q} j}\} = \{a_{\bs{k} i},b_{\bs{q} j}\} = \{b_{\bs{k} i},b_{\bs{q} j}\} \stackrel{{\bs{k}} i \neq {\bs{q}} j}{=} 0.
\end{aligned}
\end{equation}
The Hamiltonian in Eq.~\eqref{eq: BdGHamiltonianDecompositionInPHBasis} becomes a bilinear in Majorana operators. We have
\begin{equation}
\begin{aligned}
\Psi_{\bs{k} i} &= \frac{1}{\sqrt{2}} \begin{pmatrix} c_{\bs{k} i} \\ c^\dagger_{-\bs{k} i} \end{pmatrix} = \frac{1}{2\sqrt{2}} \begin{pmatrix} a_{\bs{k} i} + \mathrm{i} b_{\bs{k} i} \\ a_{\bs{k} i} - \mathrm{i} b_{\bs{k} i} \end{pmatrix} \\&=\frac{1}{\sqrt{2}} \begin{pmatrix} 1 & +\mathrm{i} \\ 1 & -\mathrm{i} \end{pmatrix} \left[\frac{1}{2} \begin{pmatrix} a_{\bs{k} i}\\ b_{\bs{k} i} \end{pmatrix} \right] \equiv S \, \Gamma_{\bs{k} i},
\end{aligned}
\end{equation}
where $S$ is the unitary matrix that transforms between the particle-hole basis $\Psi_{\bs{k} i}$ and the Majorana basis $\Gamma_{\bs{k} i}$. Comparing with Eq.~\eqref{eq: BdGHamiltonianDecompositionInPHBasis}, we deduce that the Majorana Hamiltonian is given by
\begin{equation}
\mathcal{H}_{\bs{k}}^\mathrm{M}  = S^\dagger \mathcal{H}_{\bs{k}} S,
\end{equation}
and $\mathcal{H}_{-\bs{k}}^\mathrm{M}$ has negative-energy eigenstates 
\begin{equation}
\begin{pmatrix} -v^\mathrm{M}_{\bs{k} \xi j} \\ u^\mathrm{M}_{\bs{k} \xi j} \end{pmatrix} = S^\dagger \begin{pmatrix} -v_{\bs{k} \xi j} \\ u_{\bs{k} \xi j} \end{pmatrix} = \frac{1}{\sqrt{2}} \begin{pmatrix} -v_{\bs{k} \xi j} + u_{\bs{k} \xi j} \\  \mathrm{i} v_{\bs{k} \xi j} + \mathrm{i} u_{\bs{k} \xi j} \end{pmatrix}.
\end{equation}
The inverse transformation reads
\begin{equation}
\begin{pmatrix} -v_{\bs{k} \xi j} \\ u_{\bs{k} \xi j} \end{pmatrix} = S \begin{pmatrix} -v^\mathrm{M}_{\bs{k} \xi j} \\ u^\mathrm{M}_{\bs{k} \xi j} \end{pmatrix} = \frac{1}{\sqrt{2}} \begin{pmatrix} -v^\mathrm{M}_{\bs{k} \xi j} + \mathrm{i} u^\mathrm{M}_{\bs{k} \xi j} \\ -v^\mathrm{M}_{\bs{k} \xi j} - \mathrm{i} u^\mathrm{M}_{\bs{k} \xi j} \end{pmatrix}.
\end{equation}
The constraints in Eqs.~\eqref{eq: uudaggerconstraint},~\eqref{eq: uvtransposeconstraint} become
\begin{equation}
\label{eq: uudaggerconstraintMajorana}
\begin{aligned}
&u^\mathrm{M}_{\bs{k}} (u^\mathrm{M}_{\bs{k}})^\dagger + v^\mathrm{M}_{\bs{k}} (v^\mathrm{M}_{\bs{k}})^\dagger = (v^\mathrm{M}_{\bs{k}})^\dagger v^\mathrm{M}_{\bs{k}} + (\bar{v}^\mathrm{M}_{-\bs{k}})^\dagger \bar{v}^\mathrm{M}_{-\bs{k}} \\
&= (u^\mathrm{M}_{\bs{k}})^\dagger u^\mathrm{M}_{\bs{k}} + (\bar{u}^\mathrm{M}_{-\bs{k}})^\dagger \bar{u}^\mathrm{M}_{-\bs{k}}
= \mathbb{1}, 
\end{aligned}
\end{equation}
\begin{equation}
\label{eq: uvtransposeconstraintMajorana}
\begin{aligned}
& u^\mathrm{M}_{\bs{k}} (\bar{u}^\mathrm{M}_{-\bs{k}})^\dagger + v^\mathrm{M}_{\bs{k}} (\bar{v}^\mathrm{M}_{-\bs{k}})^\dagger = (v^\mathrm{M}_{\bs{k}})^\dagger u^\mathrm{M}_{\bs{k}} + (\bar{v}^\mathrm{M}_{-\bs{k}})^\dagger \bar{u}^\mathrm{M}_{-\bs{k}} \\ & = (u^\mathrm{M}_{\bs{k}})^\dagger v^\mathrm{M}_{\bs{k}} + (\bar{u}^\mathrm{M}_{-\bs{k}})^\dagger \bar{v}^\mathrm{M}_{-\bs{k}} = 0.
\end{aligned}
\end{equation}
We are now in a position to rewrite the many-body ground state in the Majorana basis:
\begin{equation}
\label{eq: majoranagroundstate}
\begin{aligned}
\ket{\Omega}
&\propto \prod_{\bs{k}, \xi} \left[\sum_{i} \left( u_{\bs{k} \xi i} c_{\bs{k} i} - v_{\bs{k} \xi i} c^{\dagger}_{-\bs{k} i} \right) \right] \ket{0} \\
&\propto \prod_{\bs{k}, \xi} \left[\frac{1}{\sqrt{2}} \sum_{i} \left( u^\mathrm{M}_{\bs{k} \xi i} b_{\bs{k} i} - v^\mathrm{M}_{\bs{k} \xi i} a_{\bs{k} i} \right) \right] \ket{0}.
\end{aligned}
\end{equation}
\section{Quantization of the Berry phase}
\label{sec: wilsonloopappendixphs}
We here derive the quantization of the Berry phase due to particle-hole symmetry and a spectral gap. Take the $M$ dimensional occupied subspace of bands to be spanned by the eigenstates $\ket{\alpha_{\bs{k} \xi}}$, $\xi=1 \dots M$. These are related to the states considered in Sec.~\ref{sec: formalism} by
\begin{equation}
\ket{\alpha_{\bs{k} \xi}} = \alpha_{\bs{k}\xi} \ket{0}. 
\end{equation}
The action of a particle-hole symmetry $P$ with $P^2 = 1$ is $P \mathcal{H}_{\bs{k}} P^\dagger = -\mathcal{H}_{-\bs{k}}$, which implies that
\begin{equation}
\label{eq: PconstraintOnEigenstates}
P \ket{\alpha_{\bs{k} \xi}} = \sum_{\zeta=1}^M S_{\bs{k} \xi \zeta} \ket{\tilde{\alpha}_{-\bs{k} \zeta}},
\end{equation}
where $\ket{\tilde{\alpha}_{\bs{k} \xi}}$, $\xi=1 \dots M$, denotes the set of eigenstates in the empty subspace and $S_{\bs{k} \xi \zeta}$ is a unitary sewing matrix.
The Berry connection is defined as
\begin{equation}
\bs{A}_{\bs{k} \xi \zeta} = \mathrm{i} \langle \alpha_{\bs{k} \xi} |\bs{\nabla}_{\bs{k}}|\alpha_{\bs{k} \zeta} \rangle.
\end{equation}
Note that it is Hermitian, that is, it satisfies $\bar{\bs{A}}_{\bs{k} \zeta \xi} = \bs{A}_{\bs{k} \xi \zeta}$.
The Berry phase is defined as
\begin{equation}
\label{eq: BerryPhaseDefinition}
\gamma_{\bs{k}_\perp} = \int_{-\pi}^{\pi} \mathrm{d} k_{\parallel} \, \mathrm{Tr} A_{\parallel \bs{k}}. 
\end{equation}
Equation~\eqref{eq: PconstraintOnEigenstates} implies that 
\begin{equation}
\bs{A}_{\bs{k} \xi \zeta} = S_{\bs{k} \xi \tilde{\xi}} \tilde{\bs{A}}_{-\bs{k} \tilde{\xi}\tilde{\zeta}}^{\mathrm{T}} S^{\dagger}_{ \bs{k} \tilde{\zeta} \zeta} + \mathrm{i} S_{\bs{k} \xi \tilde{\zeta}} \bs{\nabla}_{\bs{k}} S^{\dagger}_{\bs{k} \tilde{\zeta} \zeta},
\end{equation}
where summation over repeated indices is implicit,
we denote by the Berry connection of the states of the empty subspace as $\tilde{\bs{A}}_{\bs{k} \xi \zeta}$, and  by $\tilde{\gamma}_{\bs{k}_\perp}$ is its Berry phase. 
We therefore have
\begin{equation}
\label{eq: PconstraintOnBerryPhaseInHigherD}
\begin{aligned}
\tilde{\gamma}_{\bs{k}_\perp} 
&= 
 \int_{-\pi}^{\pi} \mathrm{d} k_{\parallel} \, \mathrm{Tr} \tilde{A}_{\parallel \bs{k}} \\
&=  \int_{-\pi}^{\pi} \mathrm{d} k_{\parallel} \, \mathrm{Tr} A_{\parallel -\bs{k}} - \mathrm{i} \int_{-\pi}^{\pi} \mathrm{d} k_{\parallel} \, \mathrm{Tr} \left(S_{-\bs{k}} \bs{\nabla}_{-\bs{k}} S^{\dagger}_{-\bs{k}} \right) \\
&= \int_{-\pi}^{\pi} \mathrm{d} k_{\parallel} \, \mathrm{Tr} A_{\parallel (k_{\parallel},-\bs{k}_\perp)} + 2\pi \nu, \quad \nu \in \mathbb{Z} \\
&= \gamma_{-\bs{k}_\perp} \mod 2\pi,
\end{aligned}
\end{equation}
where we recognized the definition of the winding number $\nu$ of the unitary matrix $S_{\bs{k}}$ in the second term of the second line. 
Since the combination of occupied and empty bands is necessarily trivial, we have the additional constraint $\gamma_{\bs{k}_\perp} + \tilde{\gamma}_{\bs{k}_\perp} = 0$. Note that this constraint holds only in the convention where $\ket{\alpha_{\bs{k} \xi}}$ is periodic in the Brillouin zone. See Sec.~\ref{sec: nonperiodicbandappendix} for a generalization to non-periodic BdG states.
Thus, at all momenta $\bar{\bs{k}}_\perp$, where $\bar{\bs{k}}_\perp$ is equal to $-\bar{\bs{k}}_\perp$ upon the addition of reciprocal lattice vectors, we obtain
\begin{equation}
\label{eq: PconstraintOnBerryPhase}
\gamma_{\bar{\bs{k}}_\perp} = 0, \pi,
\end{equation}
and the same for $\tilde{\gamma}_{\bar{\bs{k}}_\perp}$.
\section{1D $p$-wave superconductor: Multi-band case}
We here extend the results derived for a single band in the main text to an arbitrary number of bands.
\label{sec: pwavechain}
\subsection{Particle-hole basis}
To generalize the Cooper pairing obstruction discussed in the main text to $M$ bands, which allow for non-nodal order parameters such as that of an $s$-wave superconductor, we rewrite the Berry phase in terms of the $2M \times 2M$ BdG eigenstate matrix
\begin{equation}
D_k = \begin{pmatrix} u_{k} & -v_{k} \\ -\bar{v}_{-k} & \bar{u}_{-k} \end{pmatrix}.
\end{equation}
The Berry phase is equal to half the winding number of this unitary transformation, 
\begin{equation}
\gamma = \frac{\mathrm{i}}{2} \int \mathrm{d}k \, \mathrm{Tr} \left[D^\dagger_k \partial_k D_k \right] \mod 2\pi.
\end{equation}
Particle-hole symmetry implies 
\begin{equation}
P D_k P^\dagger = D_{-k}.
\end{equation}
Let $\det D_k = e^{-\mathrm{i}\lambda_k}$. Since $P$ is antiunitary, this implies
\begin{equation}
\lambda_{-k} = - \lambda_k \mod 2\pi.
\end{equation}
At $\bar{k} = 0,\pi$ we have $\lambda_{\bar{k}} = 0,\pi$.
We then deduce
\begin{equation}
\gamma = \int_{0}^{\pi} \mathrm{d}k \, \partial_k \lambda_k \mod 2\pi = \lambda_\pi - \lambda_0 \mod 2\pi.
\end{equation}
We now prove that $\gamma = \pi$ implies that the $M \times M$ matrix $u_k$ has a zero eigenvalue at either $k = 0$ or $k = \pi$. The proof again proceeds by a reductio ad absurdum. Without loss of generality we take $\gamma = \pi$ to be realized by $\lambda_0 = 0, \lambda_\pi = \pi$, implying $\det D_0 = 1$ and $\det D_\pi = -1$. Let us assume that $u_k$ has no zero eigenvalues. It therefore has a well defined inverse, and we may reexpress $\det D_k$ as
\begin{equation}
\begin{aligned}
\det D_k &= \det u_k \det \left(\bar{u}_{-k} - \bar{v}_{-k} \frac{1}{u_k} v_k \right) \\
&= \det \left(u_k \bar{u}_{-k} \right) \det \left(1 - \frac{1}{\bar{u}_{-k}} \bar{v}_{-k} \frac{1}{u_k} v_k \right) \\
&= \det \left(u_k \bar{u}_{-k} \right) \det \left(1 + \frac{1}{\bar{u}_{-k}} \bar{v}_{-k} \bar{v}^\dagger_{-k} \frac{1}{\bar{u}^\dagger_{-k}} \right) \\
&= \det \left(u_k \bar{u}_{-k} \right) \det \left(\frac{1}{\bar{u}_{-k} u^\mathrm{T}_{-k}}\right) \\
&= 1 \text{ at } k=0,\pi.
\end{aligned}
\end{equation}
where in the third line we used that the constraint in Eq.~\eqref{eq: uvtransposeconstraint} implies
\begin{equation}
v^\dagger_{-k} \frac{1}{u^\dagger_{-k}} = - \frac{1}{\bar{u}_k} \bar{v}_k.
\end{equation}
This is however a contradiction with our earlier assertion that $\det D_\pi = -1$. We conclude that either $u_0$ or $u_\pi$ have at least one eigenvalue pinned to zero in systems with $\gamma = \pi$. This causes the Fourier transform $g_{xy,ij}$ of the Cooper pair wavefunction $g_{kij}$ in Eq.~\eqref{eq: manybodygroundstate} to have at least one eigenmode of Cooper pairs which cannot be exponentially bound together. For $\gamma = 0$ on the other hand, the Cooper pairs coming from all $M$ bands can be instantiated with an exponentially decaying separation dependence. The case where both $\det D_0 = -1$ and $\det D_\pi = -1$, corresponding to $\gamma = 0$, can be trivialized by a translational symmetry breaking perturbation such as a density wave, which flips the sign of both determinants.
\subsection{Majorana basis}
We here generalize our derivation of the Majorana charge and polarization to an arbitrary number of bands. For the moment, we discuss the more general case of $D$ dimensions. Taking the trace of the Majorana constraint in Eq.~\eqref{eq: uudaggerconstraintMajorana}, we obtain 
\begin{equation}
\begin{aligned}
\mathrm{Tr} \left[u^\mathrm{M}_{\bs{k}} (u^\mathrm{M}_{\bs{k}})^\dagger + v^\mathrm{M}_{\bs{k}} (v^\mathrm{M}_{\bs{k}})^\dagger \right] &= \mathrm{Tr} \left[(u^\mathrm{M}_{\bs{k}})^\dagger u^\mathrm{M}_{\bs{k}} + (\bar{u}^\mathrm{M}_{-\bs{k}})^\dagger \bar{u}^\mathrm{M}_{-\bs{k}} \right], \\
\mathrm{Tr} \left[v^\mathrm{M}_{\bs{k}} (v^\mathrm{M}_{\bs{k}})^\dagger \right] &= \mathrm{Tr} \left[(\bar{u}^\mathrm{M}_{-\bs{k}})^\dagger \bar{u}^\mathrm{M}_{-\bs{k}} \right], \\
\mathrm{Tr} \left[(v^\mathrm{M}_{\bs{k}})^\dagger v^\mathrm{M}_{\bs{k}} \right] &= \mathrm{Tr} \left[(u^\mathrm{M}_{-\bs{k}})^\dagger u^\mathrm{M}_{-\bs{k}} \right].
\end{aligned}
\end{equation}
We therefore find that
\begin{equation}
\sum_{\bs{k}} \mathrm{Tr} \left[(v^\mathrm{M}_{\bs{k}})^\dagger v^\mathrm{M}_{\bs{k}} \right] = \frac{M N^D}{2} = \sum_{\bs{k}} \mathrm{Tr} \left[(u^\mathrm{M}_{-\bs{k}})^\dagger u^\mathrm{M}_{-\bs{k}} \right].
\end{equation}
This implies that the total spectral weight carried by Majoranas of $a$ or $b$ type is always equal, a property that is not realized in other bases: In the particle-hole basis, for example, the total spectral weight carried by holes is zero in the case of a band insulator, and more generally takes on non-quantized and non-universal values.

In 1D real space, we may use the Majorana Wannier states $\ket{W^{\mathrm{M}}_{R, \xi}}$ of band $\xi$ and unit cell $R$ to define the Wannier functions
\begin{equation}
\braket{r, \alpha, j | W^{\mathrm{M}}_{R, \xi}} = W^{\mathrm{M}}_{R \alpha \xi j}(r) = \frac{1}{N} \sum_{k} e^{- \mathrm{i} k (R-r)} \begin{pmatrix} -v^\mathrm{M}_{k \xi j} \\ u^\mathrm{M}_{k \xi j} \end{pmatrix}_\alpha,
\end{equation}
where $\alpha = 1,2$ labels the Majorana index and $j=1\dots M$ the sublattice index. We may then compute
\begin{equation}
\sum_{r \xi j} \overline{W}^{\mathrm{M}}_{0 \alpha \xi j}(r) W^{\mathrm{M}}_{0 \alpha \xi j}(r) = \frac{1}{N} \sum_{k} \mathrm{Tr} \left[ \begin{pmatrix} v^\mathrm{M \dagger}_{k} v^\mathrm{M}_{k} \\ u^\mathrm{M \dagger}_{k} u^\mathrm{M}_{k} \end{pmatrix}_\alpha \right] = \frac{M}{2}.
\end{equation}
The total Wannier function support within a unit cell therefore splits into equal contributions to Majoranas of $a$ and $b$ type also in the multi-band case.
\section{Constraints on the Berry phase for arbitrary atomic positions}
\label{sec: nonperiodicbandappendix}
In the main text, and also so far in the Supplementary Information, we have used a convention where the BdG eigenfunctions $u_{\bs{k} \xi j}$ and $v_{\bs{k} \xi j}$ are periodic in the Brillouin zone. By this, we in effect placed all atomic orbitals that contribute to the quasiparticle spectrum at the origin of the unit cell. In general however, the BdG Hamiltonian is not periodic in the Brillouin zone, but rather satisfies 
\begin{equation}
\mathcal{H}_{\bs{k}+\bs{G} ij} = V(\bs{G})_{im} \mathcal{H}_{\bs{k} mn} V^\dagger(\bs{G})_{nj},
\end{equation}
with $\bs{G}$ a reciprocal lattice vector, and summation over repeated indices implied. This unitary transformation is determined by the atomic positions $\bs{r}_i$, $i = 1 \dots M$:
\begin{equation}
V(\bs{G})_{ij} = \begin{pmatrix} 1 & 0 \\ 0 & 1 \end{pmatrix} e^{-\mathrm{i}\bs{G} \cdot \bs{r}_i} \delta_{ij}.
\end{equation}
This leads to the physically motivated gauge choice
\begin{equation}
\label{eq: constraintonBdGvecsduetononperiodic}
\begin{pmatrix} v_{\bs{k}+\bs{G} \xi i} \\ u_{\bs{k}+\bs{G} \xi i} \end{pmatrix} = V(\bs{G})_{ij} \begin{pmatrix} v_{\bs{k} \xi j} \\ u_{\bs{k} \xi j} \end{pmatrix},
\end{equation}
which encapsulates information about the atomic positions in the boundary conditions of the BdG functions.
The quantization condition on the Berry phase, Eq.~\eqref{eq: PconstraintOnBerryPhase}, becomes modified in this convention. Crucially, the constraint $\gamma_{\bs{k}_\perp} + \tilde{\gamma}_{\bs{k}_\perp} = 0$ does not hold anymore. We can see this by noting that $\gamma_{\bs{k}_\perp} + \tilde{\gamma}_{\bs{k}_\perp}$ is insensitive to any kind of gap closing between the occupied and empty subspace. To calculate it conveniently, we may therefore use trivial eigenstates equal to the Hilbert space basis vectors, modulo phase factors stemming from the constraint~\eqref{eq: constraintonBdGvecsduetononperiodic}. We obtain
\begin{equation}
\gamma_{\bs{k}_\perp} + \tilde{\gamma}_{\bs{k}_\perp} \mod 2\pi = 4 \pi \sum_i r_{\parallel i} \mod 2\pi,
\end{equation}
where $r_{\parallel i}$ is the $i$th atom's coordinate along the direction corresponding to $k_\parallel$ in Eq.~\eqref{eq: BerryPhaseDefinition}.
On the other hand, the constraint in Eq.~\eqref{eq: PconstraintOnBerryPhaseInHigherD} due to particle-hole symmetry is still valid (importantly, the matrix $S_{\bs{k}}$ used in its derivation is always periodic in the Brillouin zone). At high-symmetry momenta $\bar{\bs{k}}_\perp = -\bar{\bs{k}}_\perp$, we have
\begin{equation}
\gamma_{\bar{\bs{k}}_\perp} \mod \pi = 2\pi \sum_i r_{\parallel i} \mod \pi.
\end{equation}
Given the atomic positions $\bs{r}_i$, the occupied subspace Berry phase therefore distinguishes between two distinct classes: Either $\gamma_{\bar{\bs{k}}_\perp} = 2\pi \sum_i r_{\parallel i} \mod 2\pi$ (the trivial case), or $\gamma_{\bar{\bs{k}}_\perp} = \pi + 2\pi \sum_i r_{\parallel i} \mod 2\pi$ (the topological case).
For a one-dimensional system, this refines our real-space picture in the Majorana basis. The $2M$ Majorana degrees of freedom originally stem from electrons at the atomic positions $r_i$, $i = 1 \dots M$. The Majorana Wannier center becomes
\begin{equation}
x_\mathrm{o} - \sum_i r_{i} \mod 1= 
\begin{cases}
0 & \text{trivial},\\
\frac{1}{2} & \text{topological}.
\end{cases}
\end{equation}
We thereby recover the generalization of the Majorana polarization discussed in the main text.
\section{Real-space picture of topological insulators}
\label{sec: insulatorappendix}
An insulator is described by a Bloch Hamiltonian of the form in Eq.~\eqref{eq: BdGHamiltonian}, with $\Delta_{\bs{k} ij} = 0$. Its ground state is given by
\begin{equation}
\ket{\Omega} = \prod_{\bs{k},\xi} \left(\sum_i u_{\bs{k} \xi i} c^{\dagger}_{\bs{k} i}\right) \ket{0},
\end{equation}
where here the $u_{\bs{k} \xi i}$ are the negative-energy eigenstates of the single-particle Hamiltonian $h_{ij}^{\bs{k}}$. When evaluated in the position basis, the ground state becomes a Slater determinant of Bloch wavefunctions $\Psi^{\bs{k} \xi}_{i}(\bs{r}) = e^{\mathrm{i} \bs{k} \bs{r}}u_{\bs{k} \xi i}$, which can be interpreted as a matrix with rows labeled by tuples $(\bs{k},\xi)$ and columns labeled by tuples $(\bs{r},i)$.
Similarly, we can span the occupied subspace by states labelled by real-space coordinates $\bs{R}$. These are the Wannier states created by the operators
\begin{equation}
\sum_{\bs{r},i} W^{\bs{R} \xi}_i(\bs{r}) c^{\dagger}_{\bs{r} i} \equiv \frac{1}{N} \sum_{\bs{k},i} e^{-\mathrm{i} \bs{k} \bs{R}} u_{\bs{k} \xi i} c^{\dagger}_{\bs{k} i},
\end{equation}
where $N$ is the number of momenta in the Brillouin zone. In position space, the ground state is then a Slater determinant of Wannier functions $W^{\bs{R} \xi}_i(\bs{r})$. This follows from the fact that the Slater determinant structure of the ground state is invariant under the Fourier transform operation that takes Bloch functions to Wannier functions (the Jacobian of the Fourier transform is unity). We can write it in a basis-independent fashion:
\begin{equation}
\ket{\Omega} = \prod_{\bs{R},\xi} \left(\sum_{\bs{r},i} W^{\bs{R} \xi}_i(\bs{r}) c^{\dagger}_{\bs{r} i}\right) \ket{0},
\end{equation}
We find that the ground state can be interpreted as a collection of independent electrons that are instantiated with a position dependence given by Wannier functions. In particular, considering for simplicity the case of a single band in one-dimension, we obtain the amplitudes
\begin{equation}
\braket{r_1 \dots r_N |\Omega} \propto A\left[W^1 (r_1) W^2 (r_2) \dots W^N (r_N)\right],
\end{equation}
where $A[\cdot]$ denotes an antisymmetrization over all positions $r_1 \dots r_N$ (compare this with the superconducting version in particle-hole basis discussed in the main text).
Now, we can maximally localize the Wannier functions $W^{\bs{R} \xi}_i(\bs{r})$ to arrive at the most local formulation of the many-body ground state of an insulator. By definition, a trivial insulator allows exponential localization, while in a topological insulator the maximally-localized functions still decay polynomially with distance as long as they are required to retain the relevant symmetries. Importantly, all one-dimensional insulators (absent spectral symmetries, which are difficult to realize in realistic systems) can be described by exponentially localized Wannier functions. This is in stark contrast to the one-dimensional topological superconductor discussed in Sec.~\ref{sec: pwavechain}: There, the wavefunction which individual Cooper pairs are instantiated in may at most decay polynomially with particle separation.
\section{Two-dimensional second-order topological superconductors}
We next discuss our results in the context of two-dimensional second-order topological superconductors (second-order TSCs)~\cite{Langbehn17,wang2018weak,SOTSC_Zhu19}. These represent novel topological phases of matter that host pointlike Majorana zero modes at the corners of two-dimensional samples. In addition to particle-hole symmetry, they require crystalline symmetries for their topological protection (otherwise, the Majorana zero modes could be adiabatically brought together along the sample edge and pairwise annihilated without a bulk gap closing). The most interesting case for our purposes is that of a second-order TSC protected by $C_4$ rotations and spinful time-reversal symmetry (hosting four Kramers pairs of Majorana corner modes in a square sample geometry), as there are so far no simple bulk topological invariants available that allow to diagnose it. In particular, all symmetry indicators and (nested) Wilson loop indices are trivial~\cite{SchindlerCorners}. We will now show that this phase can nevertheless be identified by the decay behavior of its Cooper pair wavefunctions in 2D. (Compare this with the 1D $p$-wave superconductor, which is also identifiable via the Berry phase, or the 2D chiral superconductor, which is also identifiable via the Chern number.) 

For our purposes, it is in fact enough to show that our methods can diagnose the time-reversal broken model, which in principle also can be achieved via its nonzero $C_4$ symmetry indicator invariants~\cite{benalcazar2018quantization}: In contrast to other topological invariants, the real-space decay properties we consider survive the addition of spinful time-reversal symmetry~\cite{Soluyanov2011Wannier,SoluyanovSmoothGauge} (time-reversal acts locally in real space and merely implies a Kramers doubling of all Cooper pair wavefunctions). We use the following BdG Hamiltonian to model the $C_4$-protected second-order TSC:
\begin{equation}
\begin{aligned}
H(k_x,k_y) = &(M-\cos k_x -\cos k_y) \sigma_z \tau_z + \Delta \sin k_x \sigma_z \tau_x \\&+ \Delta \sin k_y \sigma_z \tau_y + \delta (\cos k_x - \cos k_y) \sigma_y \tau_0,
\end{aligned}
\end{equation}
where $\sigma_i$ and $\tau_i$, $i=0,x,y,z$, are Pauli matrices and we abbreviate the Kronecker product $\sigma_i \otimes \tau_j$ by $\sigma_i \tau_j$. We interpret $\sigma_i$ as acting on a sublattice index while $\tau_i$ acts on particle-hole space. $C_4$ and particle-hole symmetry are represented by $C_4 = \sigma_z e^{\mathrm{i} \frac{\pi}{4} \tau_z}$ and $P = \tau_x \mathit{K}$. $H(k_x,k_y)$ is equivalent to a time-reversal broken version of the second-order topological insulator model introduced in Ref.~\onlinecite{Song17}. For $|M|<2$ and nonzero $\Delta$ and $\delta$, the model is in the second-order topological phase and exhibits four corner Majorana zero modes.

We now show that as long as we require $C_4$-symmetric Cooper pair wavefunctions (i.e., wavefunctions that are eigenfunctions of the $C_4$ operator), the nontrivial second-order topological nature of $H(k_x,k_y)$ can be diagnosed by the real-space decay behavior of these wavefunctions. To decompose the ground state into $C_4$ eigenspaces, we need to find a smooth and symmetric gauge in Bloch space. For this, we first diagonalize $H(k_x,k_y)$. Note that the term multiplying $(\cos k_x - \cos k_y)$ is proportional to the identity matrix in particle-hole space and so we can first independently diagonalize the other terms, arriving at the block-diagonal Hamiltonian
\begin{equation}
\tilde{H}(k_x,k_y) = \epsilon(k_x,k_y) \sigma_z \tau_z + (\cos k_x - \cos k_y) \sigma_y \tau_0,
\end{equation}
where we defined \begin{equation}
\epsilon(k_x,k_y) = \sqrt{3+ 2 (\cos k_x \cos k_y - \cos k_x - \cos k_y)}.
\end{equation}
\begin{widetext}
The block-diagonal form of $\tilde{H}(k_x,k_y)$ allows us to find the following smooth negative-energy eigenstates:
\begin{equation}
\label{eq: eigstatesintildespace}
\begin{aligned}
\ket{\Psi_1(k_x,k_y)}&=\frac{1}{n(k_x,k_y)}\left[\cos k_y-\cos k_x,\quad 0,\quad \mathrm{i} \sqrt{3+\cos^2 k_x+\cos^2 k_y-2(\cos k_x+\cos k_y)}+\mathrm{i} \epsilon(k_x,k_y),\quad 0\right], \\
\ket{\Psi_2(k_x,k_y)}&=\frac{1}{n(k_x,k_y)}\left[0,\quad \mathrm{i} \sqrt{3+\cos^2 k_x+\cos^2 k_y-2(\cos k_x+\cos k_y)}+\mathrm{i} \epsilon(k_x,k_y),\quad 0,\quad \cos k_x-\cos k_y\right],
\end{aligned}
\end{equation}
where we introduced the normalization factor \begin{equation}
n(k_x,k_y) = \sqrt{8 -4 (\cos k_x+\cos k_y)+\cos 2k_x+\cos 2k_y + \epsilon(k_x,k_y) \sqrt{16 -8 (\cos k_x+\cos k_y)+2(\cos 2k_x+\cos 2k_y)}}.
\end{equation}

\begin{figure}[t]
\centering
\includegraphics[width=0.8\textwidth]{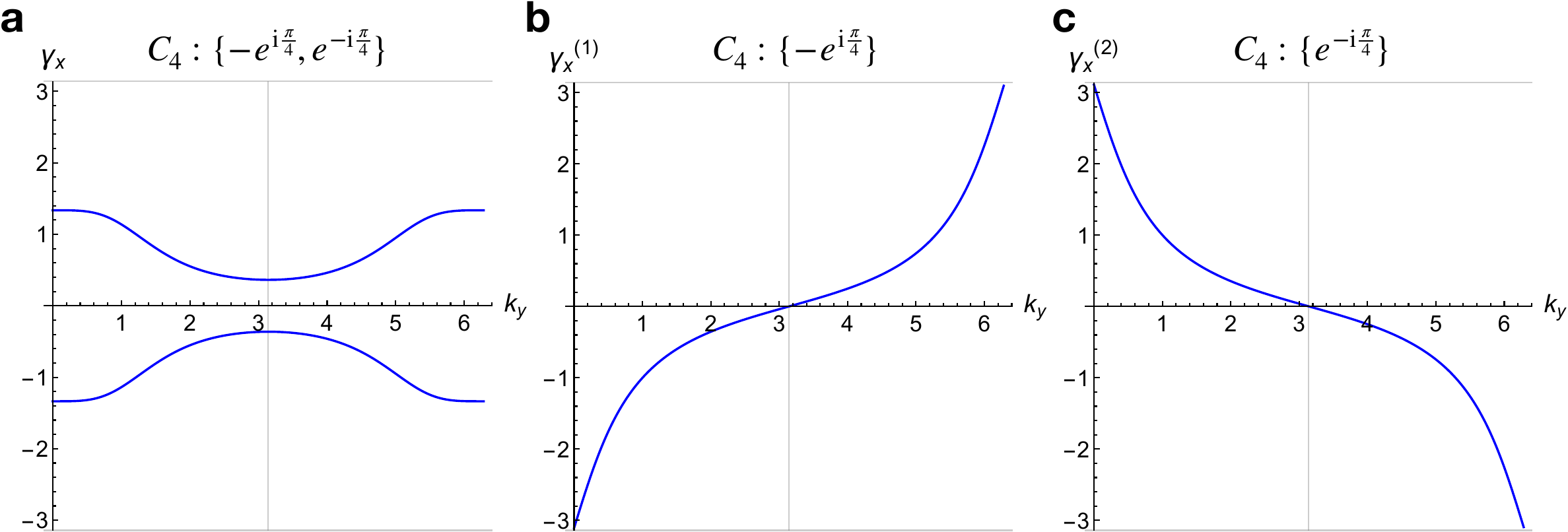}
\caption{Brillouin zone Wilson loops~\cite{Alexandradinata14,BarryFragile} in the ground state of the second-order TSC model $H(k_x,k_y)$. \textbf{a}~The total two-band Wilson loop is gapped and topologically trivial. Using a smooth $C_4$-preserving gauge, it becomes possible to calculate the Wilson loop (Berry phase) for each band separately, shown in \textbf{b}, \textbf{c}. The nontrivial second-order topology of $H(k_x,k_y)$ manifests itself via nonzero and opposite $C_4$-eigenspace Chern numbers $\pm 1$. Each $C_4$ subspace is therefore topologically equivalent to a chiral TSC, and all $C_4$-preserving Cooper pair wavefunctions [appearing in Eq.~(7) of the main text] must decay polynomially.}
\label{fig: c4sotscwilsonloops}
\end{figure}

They satisfy
\begin{equation}
\braket{\Psi_m(k_x,k_y) | \tilde{H}(k_x,k_y) | \Psi_n(k_x,k_y)} = 
\begin{pmatrix}
-\sqrt{\epsilon(k_x,k_y)^2+(\cos k_x - \cos k_y)^2} & 0 \\ 0 & -\sqrt{\epsilon(k_x,k_y)^2+(\cos k_x - \cos k_y)^2}
\end{pmatrix}_{mn},
\end{equation}
where $m,n = 1,2$, and, crucially,
\begin{equation}
\braket{\Psi_m(k_y,-k_x) | C_4 | \Psi_n(k_x,k_y)} = 
\begin{pmatrix}
-e^{\mathrm{i}\frac{\pi}{4}} & 0 \\ 0 & e^{-\mathrm{i}\frac{\pi}{4}}
\end{pmatrix}_{mn},
\end{equation}
implying that our choice of gauge is not only smooth but also $C_4$-symmetric.
\end{widetext}
Recall that $\ket{\Psi_m(k_x,k_y)}$ are the eigenstates of $\tilde{H}(k_x,k_y)$, not those of $H(k_x,k_y)$. To make a statement about the original model, we need to rotate back to the basis of $H(k_x,k_y)$. In practice, it is difficult to explicitly find a smooth unitary matrix $U(k_x,k_y)$ that achieves this (although such a choice of gauge is guaranteed to exist). We circumvent this problem by only considering gauge-invariant quantities from now on. In effect, Eq.~\eqref{eq: eigstatesintildespace} achieves a $C_4$ decomposition of the occupied subspace of $\tilde{H}(k_x,k_y)$ in the \emph{entire} Brillouin zone. It therefore allows us to calculate the Chern number \emph{per $C_4$ eigenspace}. This Chern number can be evaluated in a gauge-invariant fashion, we may therefore use any (potentially non-smooth) unitary transformation $U(k_x,k_y)$ that satisfies
\begin{equation}
U(k_x,k_y)^\dagger H(k_x,k_y) U(k_x,k_y) = \tilde{H}(k_x,k_y),
\end{equation}
and preserves $C_4$ symmetry, to rotate back to the original basis. We have obtained the transformation $U(k_x,k_y)$ by block-diagonalizing $H(k_x,k_y)$.
Figure~\ref{fig: c4sotscwilsonloops} shows the Berry phase in $x$-direction $\gamma^{(m)}_x(k_y)$, labelled by the transverse momentum $k_y$, that we obtain for the $H(k_x,k_y)$ eigenstates $U(k_x,k_y)\ket{\Psi_m(k_x,k_y)}$, $m=1,2$. From the nontrivial Berry phase winding, we can immediately read off the Chern numbers $\pm 1$ for the different $C_4$ subspaces~\cite{Taherinejad14}. The Hamiltonian $H(k_x,k_y)$ can therefore be viewed as two copies of a 2D chiral TSC, one in each $C_4$ subspace, that are coupled via a $C_4$-preserving potential. It follows that the Cooper pair wavefunctions belonging to individual $C_4$ eigenvalue subspaces are those of a chiral TSC and therefore cannot be exponentially localized, and furthermore, that the Majorana Wannier functions cannot be exponentially localized \emph{on atomic sites} (compare with Table~I from the main text). In fact, exploiting the equivalence of $H(k_x,k_y)$ with the model in Ref.~\onlinecite{Song17}, we deduce that the Majorana Wannier functions for this system will be exponentially localized not at the origin (the $1a$ Wyckoff position), but halfway along the diagonal of the $C_4$-invariant unit cell (the $1b$ Wyckoff position). Due to the $C_4$ symmetry, the two Wannier functions are pinned to the $1b$ position---the only way for functions to move symmetrically away from the $1a$ or $1b$ position is in groups of 4. This pinning ensures that Majorana modes are paired across the diagonals of the unit cell, leading to the pairing obstruction noted above. Furthermore, we can see clearly in the Majorana picture that time-reversal symmetry does not qualitatively change this picture, since although time-reversal symmetry doubles the number of Majorana Wannier functions, it also imposes the constraint that the Wannier functions must come in pairs with the same center in real space.
\bibliography{Ref-Lib}